\documentclass[pdflatex,sn-mathphys-num]{sn-jnl}

\usepackage{graphicx}%
\usepackage{multirow}%
\usepackage{amsmath,amssymb,amsfonts}%
\usepackage{amsthm}%
\usepackage{mathrsfs}%
\usepackage[title]{appendix}%
\usepackage{xcolor}%
\usepackage{textcomp}%
\usepackage{manyfoot}%
\usepackage{booktabs}%
\usepackage{algorithm}%
\usepackage{algorithmicx}%
\usepackage{algpseudocode}%
\usepackage{listings}%
\usepackage{xcolor}
\usepackage{soul}
\usepackage{lineno}
\usepackage{changes}
\usepackage{anyfontsize}

\def\fm{\hbox{$\,.\!\!^{\rm m}$}}

\def\fdg{\hbox{$\,.\!\!^\circ$}}

\def\farcs{\hbox{$\,.\!\!^{\prime\prime}$}}

\def\degr{\hbox{$^\circ$}}

\def\arcsec{\hbox{$^{\prime\prime}$}}

\def\kms{km\,s$^{-1}${}}
\def\mum{$\mu$m{}}
\DeclareRobustCommand{\ion}[2]{\textup{#1\,\textsc{\lowercase{#2}}}}
\definecolor{green2}{rgb}{0.0,0.5,0}
\begin{document}
\title[The curved jet in the young star FN~Tau]{The curved jet in the young star FN~Tau}

\author[1]{\fnm{M.~A.}~\sur{Burlak}}
\author[1]{\fnm{A.~V.}~\sur{Dodin}}
\author[1]{\fnm{A.~V.}~\sur{Zharova}}
\author[1]{\fnm{N.~P.}~\sur{Ikonnikova}}
\author[1,2]{\fnm{V.~A.}~\sur{Kiryukhina}}
\author*[1]{\fnm{S.~A.}~\sur{Lamzin}}\email{lamzin@sai.msu.ru}
\author[1,2]{\fnm{D.~A.}~\sur{Lashin}}
\author[1]{\fnm{B.~S.}~\sur{Safonov}}

\affil[1]{\orgdiv{Sternberg Astronomical Institute}, \orgname{M.V. Lomonosov Moscow State University}, \orgaddress{\street{Universitetski pr, 13}, \city{Moscow}, \postcode{119234},  \country{Russia}}}

\affil[2]{\orgdiv{Faculty of Physics}, \orgname{M.V. Lomonosov Moscow State University}, \orgaddress{\street{Leninskie gory, 1}, \city{Moscow}, \postcode{119991},  \country{Russia}}}


\abstract{In the vicinity of the young star FN~Tau, we have detected a microjet and four Herbig--Haro objects, whose positions and kinematics indicate the presence of a bipolar collimated outflow from the star --- HH\,1267. The stellar jet does not propagate rectilinearly, and we discuss the possibility that the curved shape of the jet, whose axis is inclined to the line of sight at an angle $<20^\circ$, results from the precession of the inner regions of the FN~Tau protoplanetary disk. Approximately 60 years ago, the star underwent outbursts with an amplitude of $\Delta m_{\rm pg} \sim 2^{\rm m}$ lasting several months, which we associate with the onset of the microjet.}

\keywords{stars: variable stars: T Tauri, Herbig Ae/Be --- stars: individual: DN~Tau --- ISM: jets and outflows}



\maketitle

\section{Introduction}
 \label{sect:introduct}

Classical T~Tauri stars (CTTSs) are young ($t \lesssim 10$~Myr) low-mass ($M \lesssim 2$~M$_\odot$) stars whose activity (spectral and photometric variability, line emission, etc.) is associated with magnetospheric accretion of protoplanetary disk matter \citep{BBB-1988, Hartmann-2016}. Disk accretion is accompanied by outflows from the disk in the form of disk wind and jet, which significantly affect the evolution of the angular momentum of the young star and the planet formation process \citep{Frank-2014}.

Jets are extended structures reaching up to several parsecs. They are collimated, bipolar gas flows with opening angles below $10^\circ$, ejected from the star at velocities of $\sim 300$~\kms{} \citep{Bally-2016}. Observationally, jets appear as chains of compact emission nebulae --- the so-called Herbig--Haro (HH) objects --- with the space between them filled by faintly glowing gas. Spectroscopic evidence suggests that HH objects correspond to regions behind shock fronts. In many cases, these shocks propagate at velocities nearly an order of magnitude lower than the space velocity of the HH objects themselves \citep{Dopita-2017}.

In these cases, shock waves are thought to arise when faster-moving material within the jet collides with slower gas ejected during an earlier episode \citep{Raga-1990}. Such high-velocity outflows are naturally linked to transient enhancements in the accretion rate onto the young star, which manifest as photometric brightening --- a connection directly confirmed in the case of DF~Tau \citep{Li-2001}. Since jets have been detected in fewer than 40\,\% of CTTSs \citep{Nisini-2018}, targeting stars exhibiting large-amplitude photometric variability represents a well-motivated strategy for identifying new jet sources.

In this context, the young star FN~Tau drew our attention due to its notable photometric variability during the 20th century, when its photographic magnitude ranged between $m_{\rm pg} = 14\fm7$ and $16\fm9$ \citep{Herbig-Bell-1988}. FN~Tau is projected against the dark nebula B\,209, and its parallax $\pi = 7.699 \pm 0.028$~mas \citep{Gaia-collaboration-2021} corresponds to a distance of $D \approx 130$~pc. The classification of FN~Tau as a T~Tauri star became evident after \citet{Haro-1953} detected an H$\alpha$ emission line with an equivalent width of ${\rm EW} \simeq 25$~\AA{} in its spectrum, and \citet{Goetz-1961} established that the star exhibits irregular photometric variability --- both characteristics consistent with the classical criteria for T~Tauri stars \citep{Joy-1945}.

According to \citet{Herczeg-14} and \citet{Lopez-Valdivia-2023}, the spectral type of FN~Tau is M\,3.5, corresponding to an effective temperature of $T_{\rm eff} = 3300 \pm 200$~K. Estimates of the extinction toward the star in these studies are also in close agreement: $A_{\rm V} = 1\fm15$ and $0\fm9 \pm 0\fm5$, respectively. \citet{Lopez-Valdivia-2023} further report the detection of a magnetic field in FN~Tau with a strength of $B = 1.1 \pm 0.3$~kG. Adopting a stellar luminosity of $L = 0.56$~L$_\odot$ \citep{Herczeg-14}, pre-main-sequence evolutionary models \citep{Baraffe-2015} imply a stellar mass of $M \approx 0.25$~M$_\odot$ and an age younger than $1$~Myr. The extreme youth of FN~Tau is also supported by its spectral energy distribution (SED): the dependence $\lambda F_\lambda$ vs.\ $\lambda$ remains nearly flat in the range 1--10~\mum{} \citep[Fig.\,3]{Kudo-2008}.

FN~Tau is surrounded by a protoplanetary disk that has been imaged first in scattered light at $\lambda \approx 1.6$~\mum{} using coronagraphic techniques \citep{Kudo-2008}, and later in the millimeter range with the SMA \citep{Momose-2010} and ALMA \citep{Simon-2017} interferometers. The disk is viewed nearly pole-on: the inclination of its symmetry axis to the line of sight is $i = 13^{+13}_{-8}$~degrees \citep{Dickson-Vandervelde-2025}.

The estimates of the disk accretion rate $\dot M_{\rm acc}$ for FN~Tau span a wide range, from $10^{-10}$ \citep{Lin-2023} to $10^{-8}$~M$_\odot$~yr$^{-1}$ \citep{Fang-2018}. The presence of [\ion{O}{I}] and [\ion{S}{II}] lines in the stellar spectrum \citep{Cohen-Kuhi-1979, Hirth-1997} indicates a disk wind with outflow velocities of $\sim 10$--$20$~\kms \citep{Banzatti-2019}. At the same time, \citet{Hirth-1997} identified a high-velocity component in the [\ion{O}{I}]~$\lambda 6300$~\AA{} line with a radial velocity of $V_{\rm r} = -120$~\kms{} (compared to the stellar systemic velocity of $V_{\rm r} \approx +16$~\kms), which could be interpreted as a jet signature. However, the spatial position of the emitting region coincided with the stellar position within the observational uncertainties. Thus, while \citet{Hirth-1997} provided indirect evidence for a jet in FN~Tau, our observations directly confirm this hypothesis.

The paper is organized as follows. We begin by describing our observations and the results obtained from them. We then discuss the insights gained regarding FN~Tau and its immediate surroundings. The summary and main conclusions are presented in the Conclusion.


                 
\section{Observations}
 \label{sect:observation}

To construct a historical light curve for FN~Tau, we visually estimated the star's brightness from approximately 400 photographic plates in the collection of the Sternberg Astronomical Institute, Lomonosov Moscow State University (SAI MSU) that include the star. These plates were obtained between 2 November 1951 and 26 March 1965 in a photometric system closely matching the Johnson $B$ band. Because FN~Tau lies near a dark nebula, few comparison stars were available on the plates used for our photometry. This circumstance reduced the precision of individual estimates; however, judging by the scatter of measurements obtained on the same night, the uncertainty is $\sigma_{\rm pg} < 0\fm3$. The results of the photographic photometry are available upon request.

The follow-up observations were conducted at the Caucasian Mountain Observatory (CMO) of SAI MSU \citep{Shatsky-2020}. Optical photometry of FN~Tau was obtained with the 60-cm RC600 telescope equipped with a CCD camera and $UBVR_{\rm c}I_{\rm c}$ filters in the Bessel--Cousins system \citep{Berdnikov2020}. Comparison star magnitudes were calibrated against the secondary photometric standards for the CW~Tau field from the AAVSO website\footnote{American Association of Variable Star Observers, https://www.aavso.org}. The results are presented in Table~\ref{tab:Optph} in the Appendix.

Polarimetric observations of FN~Tau in the $R_{\rm c}$ and $I_{\rm c}$ bands were obtained with the SPP speckle polarimeter mounted at the Nasmyth-2 focus of the 2.5-m telescope; the technique and reduction are described in \citet{Burlak-2026}. Speckle interferometry was performed with the same instrument on 5 November 2025. The observing strategy and data processing can be found in \citet{Safonov-2017}.

Low-resolution spectra of FN~Tau and newly identified HH objects were obtained with the Transient Double-beam Spectrograph (TDS). The instrument and reduction details are described in \citet{Potanin-2020}; here we note only that the spectral resolving power $R = \lambda/\Delta\lambda$ is $\approx 1300$ in the red channel ($0.56$--$0.74$~\mum) and $\approx 2400$ in the blue one ($0.36$--$0.56$~\mum). All wavelengths and epochs are corrected to the Solar System barycenter. The spectroscopic observation log is given in Table~\ref{tab:spectra}.

\begin{table}
\renewcommand{\tabcolsep}{0.20cm}
 \caption{Log of specrtoscopic observations} 
  \label{tab:spectra}
\begin{tabular}{cccc}
\hline\hline
BJD   &  Object  &  Exposure   &  PA      \\
2\,460\,...&           &     s    & $^\circ$ \\
\hline
953.44 & A      &  3600 & 10\\
962.52 & B, C   &  4800 & 33\\
953.49 & D      &  2400 & 45\\
983.56 & FN Tau & 1800  &  250 \\
983.54 & FN Tau & 2400  &   70 \\
984.42 & FN Tau & 1800  &  290 \\ 
984.39 & FN Tau & 1800  &  110 \\  
985.39 & FN Tau & 1800  &  350 \\
985.42 & FN Tau & 1800  &  170 \\ 
985.37 & FN Tau & 1800  &  210 \\
985.34 & FN Tau & 1800  &  30  \\
988.33 & FN Tau & 1800  &  330 \\
988.36 & FN Tau & 1800  &  150 \\
\hline
  \end{tabular}
Object labels correspond to Fig.~\ref{fig:ha}. A, B, C, D denote HH objects.  
\end{table}

On 4 October 2025, we also obtained images of the FN~Tau vicinity with a 4K$\times$4K CCD camera mounted on the 2.5-m telescope. Observations were carried out through an Halp filter ($\lambda_{\rm c} = 656$~nm, $W = 7.7$~nm; total exposure $\Delta t = 100$~min) and an Halpbc filter for the adjacent continuum ($\lambda_{\rm c} = 643$~nm, $W = 12$~nm; $\Delta t = 50$~min). The filter curves can be found at \url{https://arca.sai.msu.ru/filters?ics=NBI}.


                
\section{Results}

\subsection{Herbig--Haro objects HH~1267}  
\label{sec:jet}

\begin{figure*}
   \centering
   \includegraphics[width=\hsize]{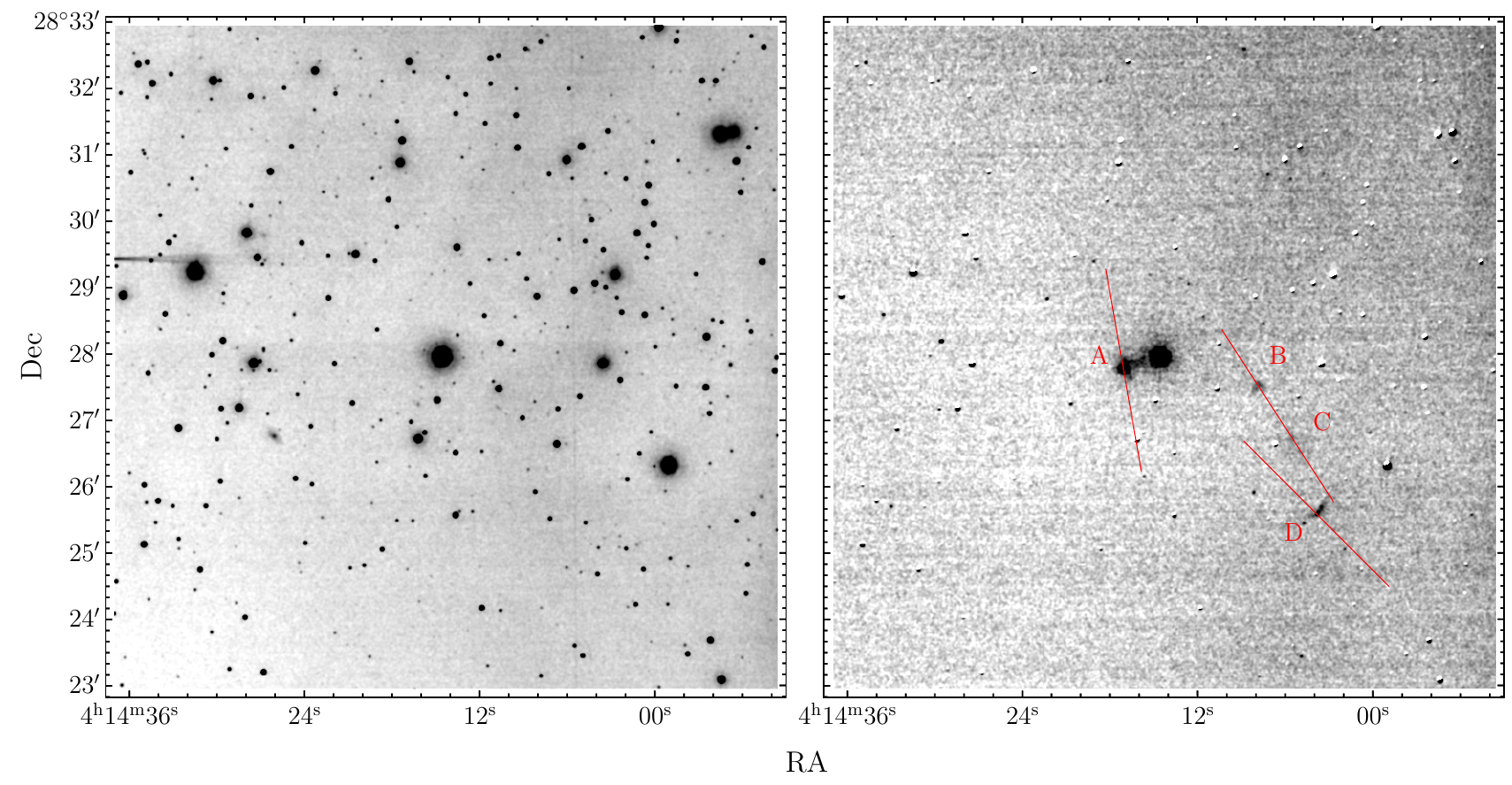}
\caption{Images of the FN~Tau vicinity in narrowband 
filters. Left: continuum in the Halpbc filter. Right:
difference image (Halp $-$ Halpbc). Red segments indicate
the slit positions used for spectroscopy of the HH 
objects labeled A, B, C, and D. 
  \label{fig:ha}
}
   \end{figure*}

The left panel of Fig.~\ref{fig:ha} shows an image of the FN~Tau vicinity obtained in the continuum near the H$\alpha$ line (Halpbc filter). FN~Tau, located in the center of the frame, lies at the northern edge of the dark nebula B\,209. The right panel displays the difference image of the same region in the Halph and Halpbc filters (see Section~\ref{sect:observation}). Emission nebulae labeled A, B, C, and D are visible on both sides of the star.

\begin{table}
\renewcommand{\tabcolsep}{0.08cm}
\caption{Parameters of the HH objects}
\label{tab:jet-knots}                     
\begin{tabular}{l c c c c c}   
\hline\hline                  
HH & $\alpha_{2000}$ & $\delta_{2000}$ & $d,$\arcsec 
& PA,$^\circ$ & $V_{\text r},$ \kms \\
\hline 
A & 04:14:17.1 & 28:27:47 & 35  & 108 & $+98 \pm 8$  \\ 
B & 04:14:07.9 & 28:27:32 & 92  & 254 & $-101 \pm 7$  \\ 
C & 04:14:05.4 & 28:26:41 & 144 & 238 & $-31 \pm 5$  \\ 
D & 04:14:03.8 & 28:25:37 & 200 & 225 & $-24 \pm 5$  \\
\hline  
FN~Tau & 04:14:14.6 & 28:27:58 &  &  &   \\
\hline 
\end{tabular}
Distance $d,$ position angle PA and\\ 
radial velocity $V_{\text r}$ are given relative to FN~Tau.
\end{table}

\begin{figure}
   \centering
  \includegraphics[width=0.55\hsize]{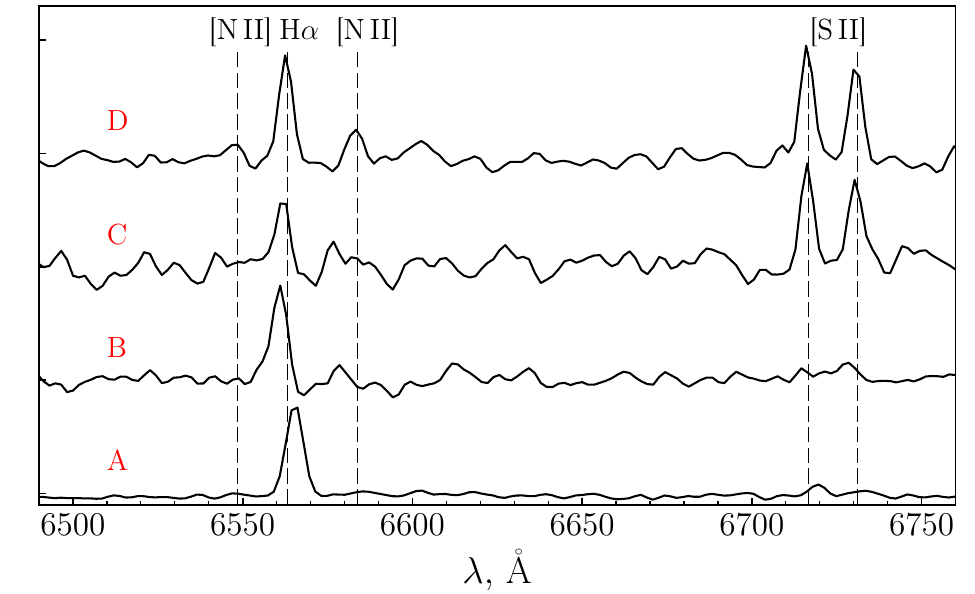}
\caption{Spectra of HH objects. Object labels correspond to Fig.~\ref{fig:ha}. Dashed lines indicate the laboratory wavelengths of selected lines. Wavelengths are corrected to the barycentric frame. 
  \label{fig:tdshh}
}
   \end{figure}

To determine the nature of these nebulae, we obtained their spectra, shown in Fig.~\ref{fig:tdshh}. The slit orientation is marked in Fig.~\ref{fig:ha} (right). The spectra exhibit H$\alpha$, [\ion{N}{II}], and [\ion{S}{II}] lines in emission shifted by several tens of \kms\ relative to FN~Tau. Table~\ref{tab:jet-knots} shows that nebula~A has $V_{\rm r} > 0$, while the three nebulae on the opposite side of the star have $V_{\rm r} < 0$.

On this basis, we identify the discovered nebulae as HH objects tracing the jet (objects B, C, and D) and counter-jet (object A) of FN~Tau. Following our arguments, Professor Bo Reipurth included the FN~Tau HH flow in the Herbig--Haro catalog \citep{Reipurth-2000}, designating it HH\,1267.


\subsection{The FN~Tau microjet} \label{sec:microjet}

As noted in the Introduction, \citet{Hirth-1997} reported a high-velocity component in the [\ion{O}{I}]~6300~\AA\ line of FN~Tau. However, the emission region's spatial position coincided with the star's position within observational uncertainties, possibly due to unfavorable spectrograph slit orientation.

\begin{figure*}
   \centering
   \includegraphics[width=\hsize]{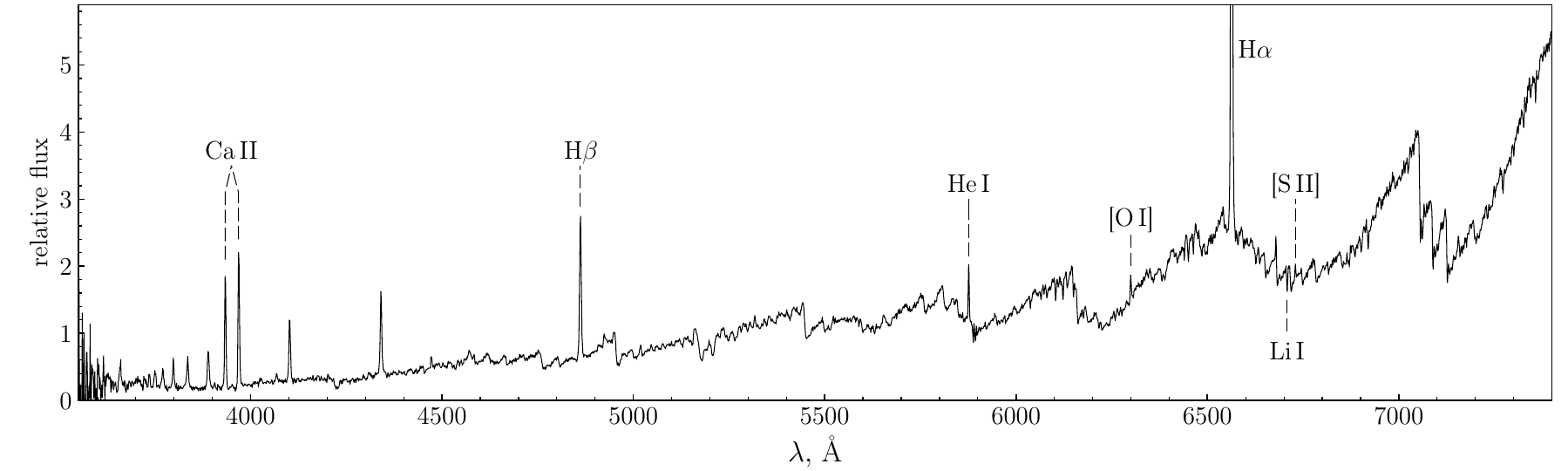}
\caption{Spectrum of FN~Tau. Laboratory wavelengths of selected lines, corrected to the barycentric frame, are indicated.
  \label{fig:tdsFN}
}
   \end{figure*}

To verify this hypothesis, we acquired a series of TDS spectra with varying spectrograph slit position angles (PA; see Table~\ref{tab:spectra}). The composite spectrum of FN~Tau is presented in Fig.~\ref{fig:tdsFN}, which shows weak  
[\ion{O}{I}]~6300~\AA\ and [\ion{S}{II}]~6731~\AA lines. However, telluric component contaminated the [\ion{O}{I}] profiles, prompting us to analyze the spatial offset $\Delta S$ of the sulfur doublet centroids relative to the stellar position as a function of slit PA. We note that \citet{Hirth-1997} did not detect a high-velocity component in the profile of [\ion{S}{II}] lines.

\begin{figure}
   \centering
   \includegraphics[width=0.5\hsize]{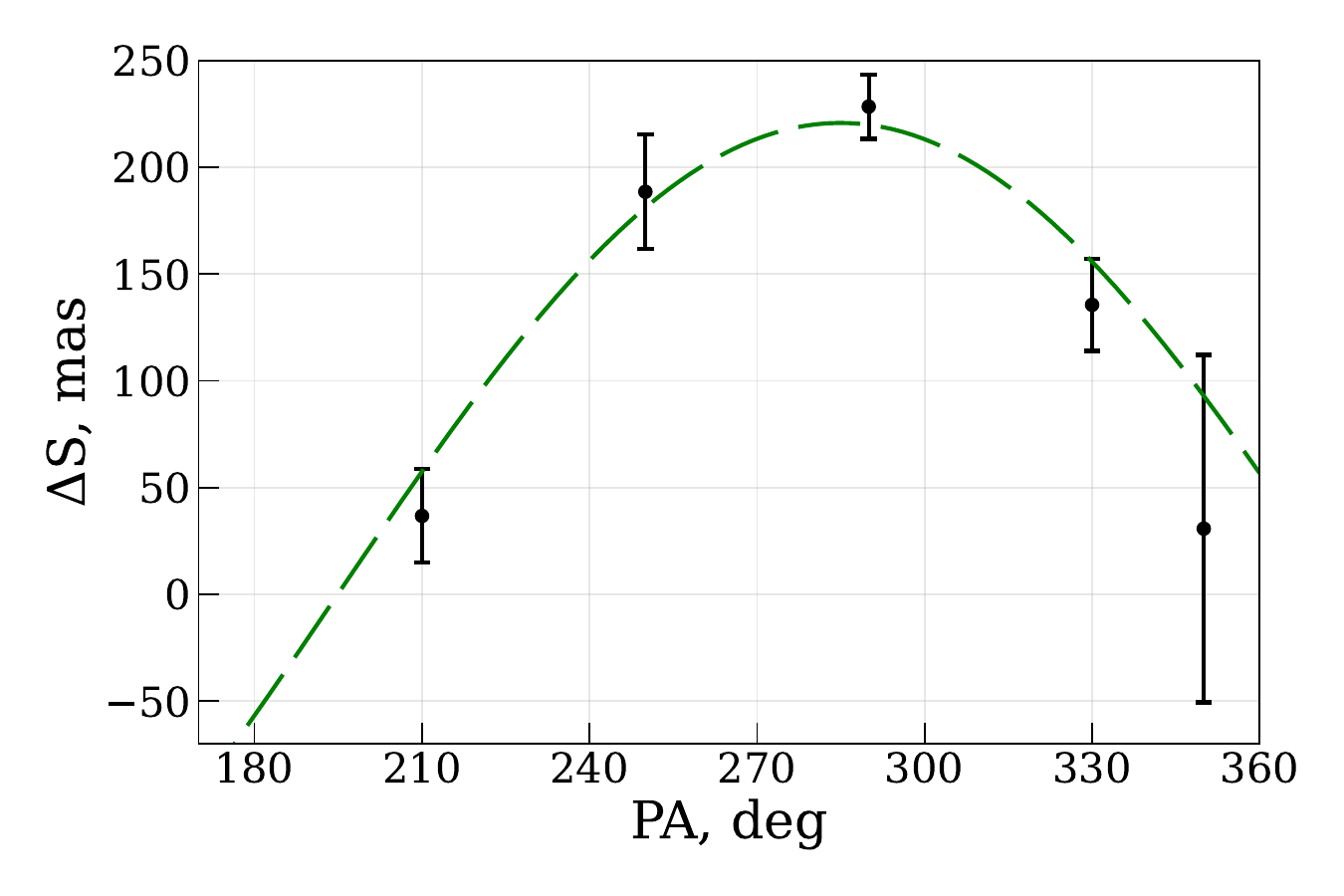}
\caption{Observed offset $\Delta S$ of [\ion{S}{II}] emission centroids relative to FN~Tau as a function of spectrograph slit PA. The dashed curve represents a cosine fit to the $\Delta S(\mathrm{PA})$ relation. See text for details. 
  \label{fig:micro}
}
   \end{figure}

The derived result is shown in Fig.~\ref{fig:micro}. Fitting the observational data with the expected theoretical relation $\Delta S = a\,\cos(\mathrm{PA} - b)$ gives $a = 221 \pm 12$~mas and $b = 285^\circ \pm 4^\circ$. Thus, the centroid of the [\ion{S}{II}] doublet emission is displaced from the stellar position by $\approx 0\farcs22$ at PA~$\approx 285^\circ$, corresponding to a projected separation $R_\bot \approx 30$~au on the plane of the sky.

From the [\ion{S}{II}]~6731~\AA\ line, we determine the radial velocity of the line-forming region, $V_{\mathrm{r}}^{\mathrm{mj}} \approx -91 \pm 3$~\kms, relative to the star. Hence, the emission originates from the approaching component of the outflow, commonly referred to as a microjet. Furthermore, Table~\ref{tab:jet-knots} shows that the PA of the microjet from FN~Tau is diametrically opposed to that of HH object A (PA$_{\mathrm{HH}}^{\mathrm{A}} \approx 108^\circ$), which traces the counter-jet.


\subsection{Results of SPP observations} 
\label{sec:spp}

Table~\ref{tab:polariz} shows that the emission from FN~Tau is only weakly polarized. The mean degrees of polarization $p$ measured in the $R_{\rm c}$ and $I_{\rm c}$ bands agree within uncertainties, both being $\approx 0.86$\%. Adopting $A_{\rm V} \approx 1$~mag toward FN~Tau (see Introduction), this polarization level is consistent with interstellar extinction \citep[\S\,18]{Bochkarev-2009}. The corresponding polarization angles are $\theta_{\rm R} = 136^\circ \pm 12^\circ$ and $\theta_{\rm I} = 110^\circ \pm 7\fdg5$, differing by less than $3\sigma$.

\begin{table}
\renewcommand{\tabcolsep}{0.08cm}
 \caption{Polarimetry in the $R_{\mathrm{c}}$ and $I_{\mathrm{c}}$ bands.}
  \label{tab:polariz}
  \begin{tabular}{l|llll|llll}
\hline
rJD &
$p_R$ & $\sigma_p$ & $\theta_R$ & $\sigma_\theta$ &
$p_I$ & $\sigma_p$ & $\theta_I$ & $\sigma_\theta$ \\
 & \% & \% & $\degr$ & $\degr$ &
\% & \% & $\degr$ & $\degr$ \\
\hline
0985.445 & 0.86 & 0.35 & 153.2 & 23.0 & 1.19 & 0.16 & 119.3 & 7.7  \\
1008.338 & 1.41 & 0.38 & 127.2 & 15.5 & 0.56 & 0.17 & 100.9 & 17.5 \\
1009.350 & 0.33 & 0.31 & 127.6 & 53.1 & 0.84 & 0.18 & 109.0 & 12.0 \\
\hline
 \end{tabular} 
${\rm rJD}={\rm JD}-2\,460\,000.$ \\
\end{table}

\begin{figure}
   \centering
\includegraphics[width=0.55\hsize,trim={0.6cm 0 0 0},clip]
{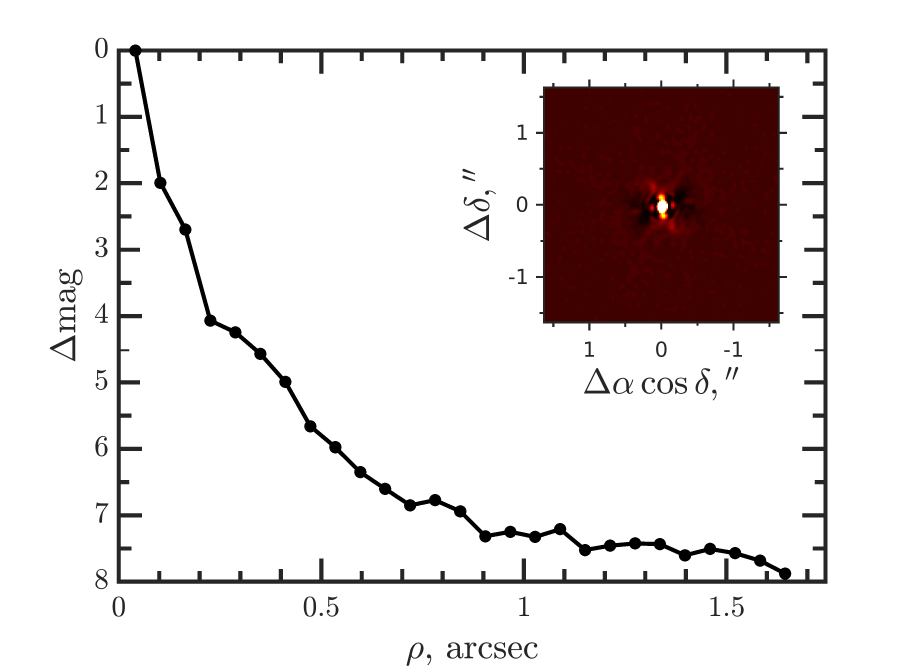}
\caption{Speckle interferometric observations of FN~Tau: dependence of detection contrast on angular separation from the central star. Insert: autocorrelation function (see text for details). 
  \label{fig:SPP_contrast}
}
   \end{figure}

Speckle interferometry in the $I_{\rm c}$ band did not detect close companions within the angular separation range $0\farcs08$--$1\farcs5$. Figure~\ref{fig:SPP_contrast} presents contrast limits for potential companions as a function of separation from FN~Tau. The insert shows the autocorrelation function (ACF) derived via the inverse Fourier transform of the mean power spectrum. A companion would manifest itself as a pair of symmetric maxima in the ACF relative to the origin \citep{Tokovinin-2010}, but no such signal is present. Weak features at $\approx 0\farcs1$ from the star are instrumental artifacts caused by optical aberrations and mount jitter, as confirmed by similar patterns in other targets observed during the same night.

\subsection{Historical light curve of FN~Tau}  
\label{sec:lcurve}

\begin{figure}
   \centering
   \includegraphics[width=\hsize]{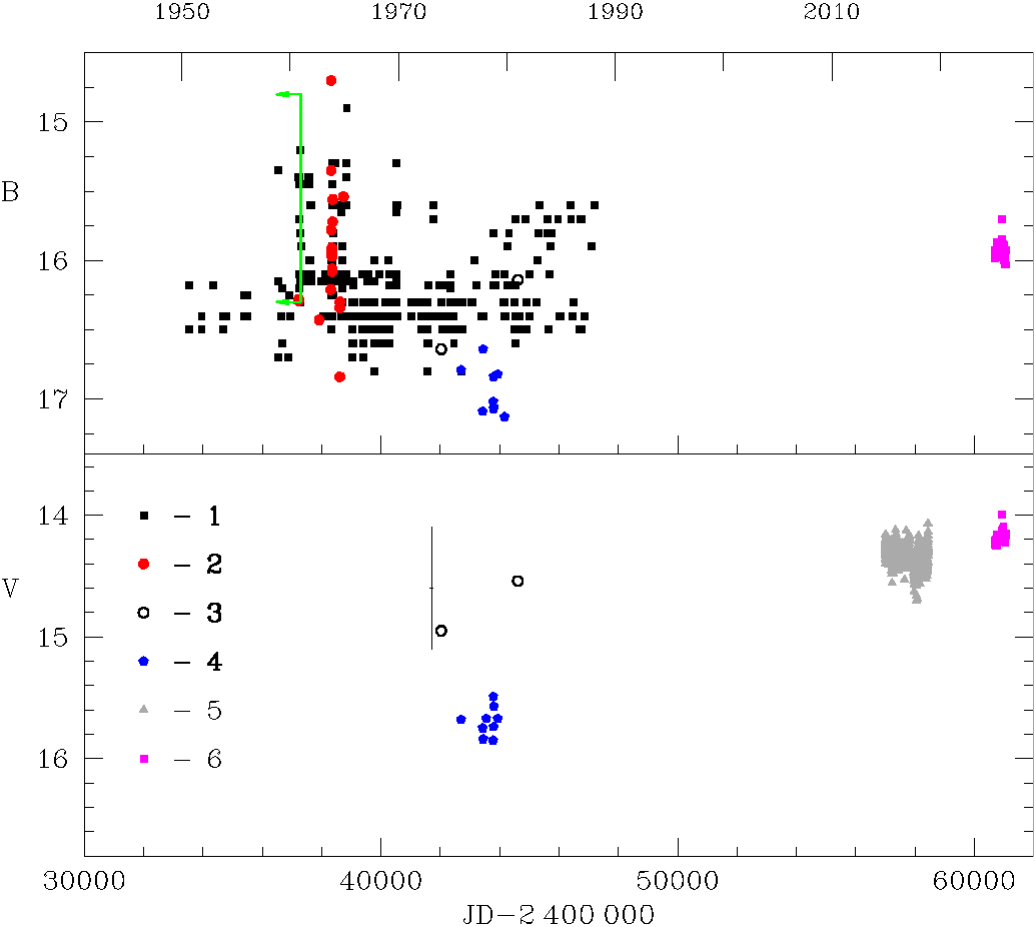}
\caption{Historical $B$- and $V$-band light curves of FN~Tau (upper and lower panels, respectively). Different symbols and colors represent data from: (1) our photographic measurements; (2) \citet{Badalian-1969}; (3) \citet{Rydgren-1976,Rydgren-1982}; (4) \citet{Nurmanova-1983}; (5) ASAS-SN survey \citep{Kochanek-2017}; (6) our CCD observations. Vertical lines indicate measurements from \citet{Goetz-1961} (top panel) and \citet{Romano-1975} (bottom panel); details are provided in the text. 
  \label{fig:lcurve}
}
   \end{figure}

Figure~\ref{fig:lcurve} shows the historical $B$- and $V$-band light curve of FN~Tau, compiled from our observations and archival data from \citet{Badalian-1969}, \citet{Rydgren-1976,Rydgren-1982}, \citet{Nurmanova-1983}, and the ASAS-SN survey \citep{Kochanek-2017}. Additionally, the upper panel includes a green square bracket with arrows indicating that \citet{Goetz-1961} reports only a brightness range ($16\fm3$--$14\fm8$) without specifying the corresponding time interval. Furthermore, \citet{Romano-1975} notes that the $V$-band brightness varied between $15\fm1$ and $14\fm1$ from 25 November 1972 to 20 March 1973, but individual 
photometric measurements are not listed in that work.

A notable feature of the light curve is the significant brightening of FN~Tau in the $B$ band during the first half of the 1960s. Apparently, at least two outbursts of comparable amplitude can be identified, each lasting several months and separated by an interval of approximately two months. 

            
\section{Discussion}
\label{sec:discuss}

The FN~Tau jet and counter-jet exhibit pronounced asymmetry: three HH objects (B, C, and D) are detected in the jet, while only one (object A) is present in the counter-jet (Fig.~\ref{fig:ha}). The outflow asymmetry occurs in approximately 50\% of CTTSs, with jets and counter-jets showing differences in both the morphology and physical properties of their associated HH objects \citep{Hirth-1994}. These differences are attributed to environmental inhomogeneities on opposite sides of the protoplanetary disk and/or to the asymmetric activity of the ``central engine''{} \citep{Bally-2016}.

In any case, the absence of a detectable counter-{\it microjet} can be straightforwardly explained: FN~Tau's protoplanetary disk has a radius of $\sim 2\arcsec$ \citep{Kudo-2008}, and the system is viewed at a low inclination $i \lesssim 20^\circ$ relative to the disk axis (see Introduction). Consequently, the disk should obscure any counter-microjet located at a separation ($\sim 0\farcs2$) similar to the observed microjet.

The nearly pole-on orientation of the disk-star system likely accounts for the low polarization observed in FN~Tau. In CTTSs, substantial polarization typically results from scattering of stellar light by the protoplanetary disk \citep{Grinin-1988} and/or by dust grains carried outward in the disk wind \citep{Dodin-2019}. For an axially symmetric disk and dusty wind viewed nearly pole-on, the net polarization produced by these scattering mechanisms should be small.

A non-trivial feature of the FN~Tau jet is that the position angle of the HH objects tracing the jet (objects B, C, and D) decreases with increasing distance from the star (see Table~\ref{tab:jet-knots}). This means that the chain of these objects does not form a straight line; i.e. the jet bends clockwise with increasing distance from the star.

This morphology may result from interactions between 
the jet and density inhomogeneities in the ambient interstellar medium, such as remnants of the protostellar cloud \citep{Raga-book-2020}. A similar mechanism could explain the $\sim 90^\circ$ deflection observed in the DF~Tau jet (see Fig.~1 in \citealt{Dodin-2025}).

An alternative explanation for the observed curvature of the FN~Tau jet is also possible. In the vicinity of some CTTSs, chains of HH objects are elongated along an S-shaped (spiral) curve, forming so-called wiggling jets, as in the cases of V1331~Cyg \citep{Mundt-1998} and Haro~6-10 \citep{Devine-1999}. Such spiral jet morphologies are commonly attributed to precession of the inner regions of the accretion disk \citep{Bally-2016}.

For FN~Tau, such a spiral should be viewed nearly along its axis. To illustrate how it should appear, we considered a simple model. We assume that the jet is a thin stream that, due to precession with a period $P_{\rm pr}$, traces out a cone with an opening angle $2\beta$ around an axis inclined to the line of sight at an angle $i$.

Section~17.1 of the monograph by \citet{Raga-book-2020} provides formulas that describe the observed trajectory of a precessing jet in the case where its velocity $V_{\rm j}$ does not change with time, the so-called ballistic regime. However, in some CTTSs, the jet velocity decreases significantly with increasing distance from the star due to interaction with the interstellar medium (see, e.g. \citealt{Melnikov-2009,Devine-1997}).

\begin{figure}
   \centering
\includegraphics[width=0.5\hsize]{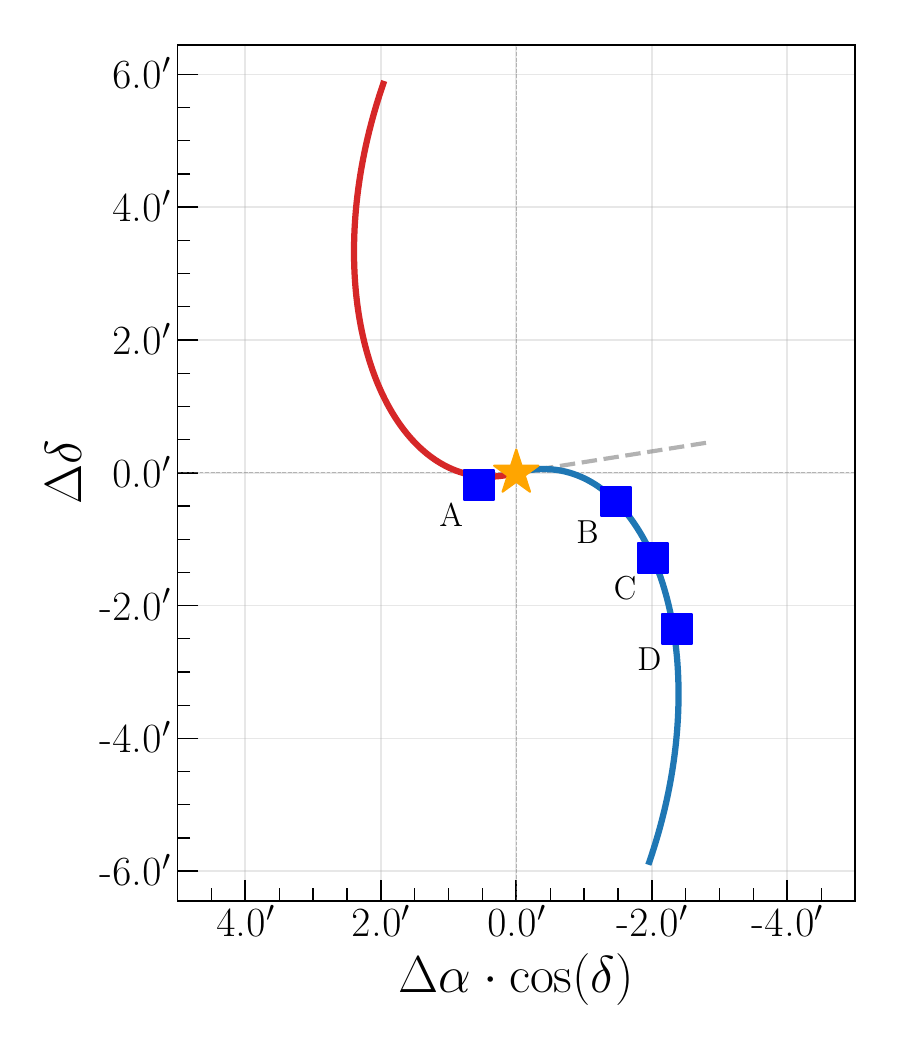}
\caption{Comparison of the observed positions of HH objects (blue squares) with model calculations. The blue curve represents the approaching spiral, the red curve the receding one. HH object labels are the same as in Fig.~\ref{fig:ha}. The dashed line indicates the direction to the microjet. See text for details. 
  \label{fig:spiral}
}
   \end{figure}

To account for this effect, we assumed that the jet velocity evolves with time according to
\begin{equation}
 \frac{{\rm d} V_{\rm j}}{{\rm d} t} =
  -\frac{k_{\rm j}}{P_{\rm pr}}  V_{\rm j}^2,
 \label{eq:V-vs-t}
\end{equation}
where $k_{\rm j}$ is a model parameter. Using the radial velocities and coordinates of HH objects A, B, C, D, and the microjet, we fitted the parameters $i$, $\beta$, $P_{\rm pr}$, $k_{\rm j}$, and the initial jet velocity $V_{\rm j}^0$. The model provides the best match to the data (Fig.~\ref{fig:spiral}) for $i \approx 14^\circ$, $\beta \approx 28^\circ$, $k_{\rm j} \approx 0.18$~s~km$^{-1}$ and $V_{\rm j}^0 \approx 110$~\kms.

Regarding the precession period $P_{\rm pr}$, the available data only allow us to state that it certainly exceeds $5\times 10^4$~yr. As seen in Fig.~\ref{fig:spiral}, this uncertainty arises because the ages of the HH objects, or more precisely their so-called dynamical timescales $t_{\rm dyn}$, are much shorter than the precession period. Nevertheless, varying $P_{\rm pr}$ from $5\times 10^4$ to $2\times 10^5$~yr does not produce a noticeable change in the morphology of the spiral arm on the scale of the figure.
    
Although our model is largely illustrative, the resulting dynamical age estimates for the HH objects are probably realistic, ranging from $t_{\rm dyn} \sim 10^3$~yr for object A to $\sim 10^4$~yr for object D. However, the microjet age can be estimated independently of the model, using the values of $R_\bot$ and $V_{\rm r}^{\rm mj}$ derived in Section~\ref{sec:microjet}, and the relation
\begin{equation}
 t_{\rm dyn}^{\rm mj} \sim  
\frac{R_\bot}{V_{\rm r}^{\rm mj}\,{\rm tan}\,i} \sim 10\,\, 
 {\hbox{yr}}.
 \label{eq:tdyn}
\end{equation}

This dynamical timescale implies that the microjet likely originated during outburst activity in FN~Tau approximately 60 years ago (see Section~\ref{sec:lcurve}).\footnote{We note that the $R_\bot$ value used in this estimate represents the offset of the [\ion{S}{II}]~6731~\AA\ line emission centroid relative to the star; the actual extent of the microjet may be significantly larger.} Similar results were obtained, for example, for such objects as Z CMa \citep{2010ApJ...720L.119W} and PV Cep \citep{2025A&A...693A..79M}.

\citet{Li-bubble-Taurus-2015} detected molecular gas motion in the vicinity of FN~Tau with a radial velocity $V_{\rm r} \sim 3$~km~s$^{-1}$ --- the CO molecular outflow TMO\_03 (see Fig.~4 in that work). The receding molecular gas is concentrated in a region with an extent of $\approx 0.5$~pc, located on both sides of the star along ${\rm PA} \approx \pm 60^\circ$. Meanwhile, the approaching gas, occupying a region of similar extent, is elongated in a nearly perpendicular direction, with FN~Tau situated approximately $2^\prime$ to the northwest of the intersection of these two regions. According to \citet{Li-bubble-Taurus-2015}, the dynamical timescale $t_{\rm dyn}$ of the approaching region is $1.3\times 10^5$~yr, while that of the receding region is $1.9\times 10^5$~yr.

It is difficult to explain the observed morphology of the TMO\_03 molecular outflows. However, comparing this morphology with the positions of much younger HH objects, we can conclude that the direction of mass outflow from the FN~Tau vicinity changes substantially over time. Such phenomena in young stars are typically attributed to precession of the inner disk regions, induced by gravitational interactions with a companion on a highly eccentric orbit \citep{Bally-2016,Fendt-2022}. This raises the question: is FN~Tau a binary system?

Attempts to detect a companion to FN~Tau have so far been unsuccessful. Speckle interferometry provides only upper limits on the magnitude difference $\Delta m$ between a hypothetical companion and the primary star as a function of their angular separation at 0.9~\mum\ (Fig.~\ref{fig:SPP_contrast}) and 4.8~\mum\ \citep{Sanghi-2024}. Unfortunately, translating these $\Delta m$ limits into a companion mass upper bound via the approach of \citet{Burlak-2025} is precluded by the uncertain evolutionary tracks and isochrones for low-mass stars at ages $t \lesssim 1$~Myr \citep{Baraffe-2015,Zallio-2026}. We also note that the astrometric solution for FN~Tau has ${\rm RUWE} = 1.32$ \citep{Gaia-collaboration-2021}, i.e., it is relatively small --- a property atypical for binary stars.

Radial velocity monitoring of FN~Tau has also failed to reveal a companion. We uniformly processed 25 archival high-resolution spectra ($R > 50\,000$) obtained with the ESPaDOnS\footnote{https://www.cadc-ccda.hia-iha.nrc-cnrc.gc.ca/en/cfht/} and UVES\footnote{http://archive.eso.org/cms/eso-data.html} spectrographs between 2013 and 2024, and found a radial velocity scatter of $\sigma_{\rm V} \approx 0.2$~\kms. This is consistent with the measurements of \citet{SDSS-Collaboration-2025}, who reported $\sigma_{\rm V} \approx 0.09$~\kms\ from their data covering approximately the same period. Such small \textit{upper limits} on $V_{\rm r}$ variations preclude any firm conclusions about binarity, as variability at this level can arise from temperature inhomogeneities (spots) on the stellar surface. In this context, we note that although FN~Tau is viewed nearly pole-on, the broadening of photospheric lines due to axial rotation is relatively large: $V_{\rm eq}\sin i = 6.4 \pm 0.4$~\kms\ \citep{Kounkel-2019}.

In summary, and given that the mass of FN~Tau is $M \approx 0.25$~$M_\odot$, we suggest that if the star has a companion, it is most likely either a brown dwarf or a giant planet.


\section{Conclusion}
\label{sec:conclusion}

The positions and kinematics of the Herbig-Haro objects HH\,1267 and the microjet we discovered indicate that a bipolar collimated outflow of matter occurs in FN~Tau, with the stellar jet not moving in a straight line. It is possible that the jet changes direction as a result of interactions with the interstellar medium, but it is more likely that the jet's curved shape results from precession of the inner regions of FN~Tau's accretion disk. This interpretation is supported not only by the ability to reproduce the observed jet shape in the spiral jet model, but also by the difference between the shape of the HH\,1267 flow and the morphology of older molecular CO outflows.

If the jet's precession is caused by the presence of a companion to FN~Tau, then, based on the available data, this companion is either a brown dwarf or a giant planet.

The formation of jets from CTTSs is associated with 
the dynamical interaction of the inner regions 
of the accretion disk with the global magnetic
field of the star, but the details of this 
process remain the subject of debate --- 
see, e.g., \citet{Ferreira-2006, Beskin-2023}
and references therein. The accumulation of observational data on the physical parameters and 
morphology of jets from CTTSs with different characteristics should help 
constrain these models.

In the Introduction, we noted that the search for jets in stars with large 
amplitudes of photometric variability is well motivated, as confirmed in 
the case of FN~Tau: the emergence of its microjet is likely associated 
with outburst activity that occurred approximately 60 years ago.

\backmatter

\bmhead{Acknowledgements}
We thank the staff of the CMO SAI MSU for the help
with observations, the anonymous referee for useful remarks, and Prof. Bo Reipurth for including the discovered HH-flow in the general catalog of objects of this type and assigning it the number HH\,1267. We also gratefully acknowledge the use of the ASAS-SN, SIMBAD (CDS, Strasbourg, France), Astrophysics Data System (NASA, USA), ESO Data, Canadian Astronomy Data Centre, and AAVSO databases in this work. The study was conducted under the state assignment of Lomonosov Moscow State University. Scientific equipment used in this study was bought partially through the M.~V.~Lomonosov Moscow State University Program of Development.

\begin{appendices}
\section{Optical photometry}
\label{sect:addition}
\begin{table*}[h!]
\renewcommand{\tabcolsep}{0.2cm}
 \caption{Results of FN~Tau's optical photometry}
  \label{tab:Optph}
  \begin{footnotesize}
  \begin{tabular}{lllll|lllll}
\hline
JD-2\,46.\,... & $B$ &$\sigma_B$ &$V$ &$\sigma_V$ &
JD-2\,46.\,... &$B$ &$\sigma_B$ &$V$ &$\sigma_V$ \\
\hline
0703.348  & 15.928  & 0.01  & 14.206 & 0.01 &
0975.523  & 15.930  & 0.01  & 14.143 & 0.01 \\
0706.218  & 15.982  & 0.01  & 14.242 & 0.01 &
0979.475  & 15.940  & 0.02  & 14.139 & 0.01 \\
0708.266  & 15.939  & 0.01  & 14.233 & 0.01 &
0981.431  & 15.900  & 0.01  & 14.118 & 0.02 \\
0710.247  & 15.930  & 0.02  & 14.218 & 0.01 &
0982.382  & 15.881  & 0.01  & 14.091 & 0.01 \\
0715.198  & 15.986  & 0.01  & 14.229 & 0.01 &
0989.331  & 15.953  & 0.01  & 14.101 & 0.01 \\
0717.203  & 15.954  & 0.02  & 14.248 & 0.01 &
0990.439  & 15.940  & 0.01  & 14.143 & 0.02 \\
0719.219  & 15.958  & 0.02  & 14.212 & 0.01 &
0996.389  & 15.932  & 0.01 & 14.147  & 0.01 \\
0729.209  & 15.941  & 0.01  & 14.199 & 0.01 &
0998.463  & 15.945  & 0.02 & 14.181  & 0.01 \\
0735.250  & 15.958  & 0.01  & 14.212 & 0.01 &
1001.279  & 15.984  & 0.01 & 14.218  & 0.01 \\
0739.181  & 15.979  & 0.01  & 14.227 & 0.02 &
1002.457  & 16.003  & 0.01 & 14.206  & 0.01 \\
0743.186  & 15.867  & 0.01  & 14.157 & 0.01 &
1003.308  & 15.980  & 0.01 & 14.196  & 0.01 \\
0745.184  & 15.943  & 0.03  & 14.188 & 0.01 &
1004.218  & 15.965  & 0.01 & 14.160  & 0.01 \\
0747.236  & 15.951  & 0.01  & 14.210 & 0.01 &
1007.264  & 15.889  & 0.01 & 14.148  & 0.01 \\
0750.196  & 15.947  & 0.01  & 14.214 & 0.01 &
1008.357  & 15.948  & 0.01 & 14.174  & 0.01 \\
0759.234  & 15.965  & 0.01  & 14.233 & 0.01 &
1009.261  & 15.965  & 0.01 & 14.194  & 0.01 \\
0765.225  & 15.932  & 0.01  & 14.254 & 0.05 &
1026.397  & 16.027  & 0.01 & 14.212  & 0.01 \\
0934.492  & 15.704  & 0.01  & 13.993 & 0.01 &
1035.352  & 15.998  & 0.01 & 14.175  & 0.01 \\
0940.535  & 15.846  & 0.01  & 14.124 & 0.01 &
1048.400  & 15.983  & 0.01 & 14.224  & 0.01 \\
0957.341  & 15.882  & 0.02  & 14.105 & 0.01 &
1053.277  & 15.955  & 0.01 & 14.166  & 0.01 \\
0962.524  & 15.922  & 0.01  & 14.150 & 0.02 &
1060.349  & 16.030  & 0.01 & 14.166  & 0.01 \\
0966.410  & 15.899  & 0.01  & 14.112 & 0.01 &
1079.215  & 15.926  & 0.01 & 14.152  & 0.01 \\
0968.370  & 15.938  & 0.01  & 14.137 & 0.01 &
          &         &       &        &      \\
\hline
 \end{tabular} \\
 \end{footnotesize}
\end{table*}
\end{appendices}

\bibliography{fntau}

\end{document}